\documentclass[aps,prd, superscriptaddress,showpacs]{revtex4}
\usepackage{amsmath,amssymb}

\usepackage{graphicx}

\newcommand{\beq}{\begin{equation}}
\newcommand{\eeq}{\end{equation}}
\newcommand{\be}{\begin{equation}}
\newcommand{\ee}{\end{equation}}

\begin{document}
\title{ Physics of Trans-Planckian Gravity}

\author{Gia Dvali}
\email{georgi.dvali@cern.ch}
\affiliation{Arnold Sommerfeld Center, Ludwig-Maximilians-University, Theresienstr. 37, 80333 Muenchen, Germany}
\affiliation{Max-Planck-Institut f\"ur Physik, 
F\"ohringer Ring 6, 80805 M\"unchen, Germany}
\affiliation{CERN,
Theory Division, 
1211 Geneva 23, Switzerland}
\affiliation{CCPP,
Department of Physics, New York University,
4 Washington Place, New York, NY 10003, USA}
\author{Sarah Folkerts}
\email{sarah.folkerts@physik.uni-muenchen.de} 
\affiliation{Arnold Sommerfeld Center, Ludwig-Maximilians-University, Theresienstr. 37, 80333 Muenchen, Germany}
\author{Cristiano Germani}
\email{cristiano.germani@physik.uni-muenchen.de}
\affiliation{Arnold Sommerfeld Center, Ludwig-Maximilians-University, Theresienstr. 37, 80333 Muenchen, Germany}



\begin{abstract}
We study the field theoretical description of a generic theory of gravity flowing to Einstein General Relativity in IR. We prove that, if ghost-free, in the weakly coupled regime such a theory can never become weaker than General Relativity. Using this fact, as a byproduct, we suggest that in a ghost-free theory of gravity
trans-Planckian propagating quantum degrees of freedom cannot exist. The only physical meaning of a trans-Planckian pole is the one of a classical state (Black Hole) which is described by the light IR quantum degrees of freedom and gives exponentially-suppressed 
contributions to virtual processes.  In this picture Einstein gravity is UV self-complete, although not  Wilsonian,
and sub-Planckian distances are unobservable in any healthy theory of gravity.  We then finally show that this UV/IR correspondence puts a severe constraint on any attempt of conventional Wilsonian UV-completion of trans-Planckian gravity.  Specifically,  there is no well-defined energy domain in which  gravity  could become asymptotically weak or safe.

\end{abstract}


\maketitle


\section{Introduction}
 Einsteinian gravity, described by the Einstein-Hilbert action, 
 \begin{equation}
 S_{EH} \, = \,  \int \, d^4x \, \sqrt{-g} \, {1 \over 16\pi \, G_N} \,  R \, ,
 \label{hilbert}
 \end{equation}
  is a unique theory which propagates a single massless spin-2 graviton, $h_{\mu\nu}$, and no other degrees 
  of freedom. 
  
  The reduced Planck mass, $M_P\, \equiv \sqrt{1 \over 16\pi\, G_N  \, }\sim\, 10^{18}$GeV, 
  and the corresponding Planck length, $L_P\, \equiv \, M_P^{-1}\, \sim \, 10^{-32}$cm, play a central role in general relativity (GR).  For example, from the field theoretical point of view, $M_P$ sets the interaction strength of the canonically-normalized graviton expanded around Minkowski,
  \begin{equation}
  {1 \over M_P} \, h_{\mu\nu}\, T^{\mu\nu} \, .
  \label{gravitoncoupling}
  \end{equation}
  Here, $T_{\mu\nu}$ is an arbitrary conserved energy-momentum source. A very special property of gravity is that also self-interactions are regulated by the coupling (\ref{gravitoncoupling}), where in this case $T_{\mu\nu}$ is the energy-momentum tensor of the graviton evaluated to a given non-linear order in $h_{\mu\nu}$. 
 
In Einsteinian gravity all energy-momentum sources universally couple to gravity. 
At linearized level, one can thus define an effective dimensionless parameter describing the strength of the gravitational interaction for any elementary process of characteristic momentum transfer  $p$,  
\begin{equation}
\alpha_{grav}(p^2) \, \equiv \, 16 
\pi \, G_N\, p^2 \, . 
\label{alphagravity}
\end{equation}
Where here, and throughout the paper, we assumed only asymptotically flat spaces so that gravity may be expanded in terms of linear gravitons up to the strong coupling scale of the theory. Note that in this way one can construct gauge invariant (e.g. diffeomorphism invariant with respect to the background metric) global and local operators such as the S-Matrix and/or the scattering amplitude $A(p)$ of a scattering process prepared at spatial infinity.

The parameterization of \eqref{alphagravity} the gravitational strength clearly shows why gravity is weak in low energy processes, 
$p \, \ll \, M_P$ (or IR). In this way Einsteinian gravity admits a universal strong coupling scale,  $M_P$.   The above energy-dependence of the effective gravitational coupling is the source of the non-renormalizability of Einstein gravity and the reason why gravitational amplitudes violate perturbative unitarity  above the scale $M_P$. 
  
 Notice that the coupling parameterization is equally applicable to extensions of Einstein gravity in which gravity is mediated by additional degrees of freedom, but which still obey the Strong Equivalence Principle \cite{Wald}.

For our purposes, it is useful to parameterize the notion of the gravitational strength and of its UV-completion by the behavior of gravitational scattering amplitudes.

Consider a scattering on asymptotically flat space among two conserved {\it external} sources  \footnote{ For a brief discussion on the notion of external sources in gravity see appendix \ref{A}.} $T_{\mu\nu}$ and $\tau_{\mu\nu}$ with characteristic momentum-transfer $p$. Throughout the paper we will only be interested in sources that do not violate null energy conditions.
the scattering amplitude can be written as
\begin{equation}                                                                                                                                                                                                                           
A(p) \, = {\alpha_{grav}(p) \over (p^2)^2} \left(T_{\mu\nu}\tau^{\mu\nu} \, + \, b(p) \, T^{\mu}{}_{\mu} \tau^{\nu}{}_{\nu}\right) \, .                                                                                                               
\label{Ap}                                                                                                                                                                                                                                 
\end{equation}                                                                                                                                                                                                                             
Here, $b(p)$ may in general depend on $p$ non-trivially.
At this point, $\alpha_{grav}(p)$ is just a useful parametrization of the gravitational strength.  Notice that in any theory in which gravitational interactions are mediated by spin-2 states, the parameter $b(p)$ is generically of order one.  However, it can be larger  if contributions from spin-0 dominate.  Such a case can be easily incorporated in our discussions, but is not of our primary interest. Moreover, the dependence on $b(p)$ can be eliminated by taking at least one of the sources to be relativistic  (say $\tau_{\mu}^{\mu}\, = \, 0$).
                                                                                                                                
Universally, the scale of strong gravity can be defined as the lowest energy scale $M_*$ for which   
  \begin{equation}
  \alpha_{grav}(p\equiv M_*) \, = \, 1\, .
  \label{strongscale}
  \end{equation}
  
In pure Einstein gravity $M_* \, = \, M_P$, but, in general, $M_*$ can be arbitrarily lower though {\it never}  higher \cite{brustein-dvali-veneziano}, as we will discuss. In any given theory, we refer to the region of energies $p \, \gg \, M_*$ as the trans-Planckian region (or UV) and to the corresponding length scales $L \, \ll \, L_* \, \equiv \, M_*^{-1}$ as sub-Planckian distances. 
  
In quantum field theory one always describes physics at any given length scale in terms of propagating quantum degrees of freedom.  In this sense, all existing states of the theory (including the classical ones)  are in principle accounted for as states of  degrees of freedom which are propagating at the length scales of interest.  Of course, when one moves from scale to scale,  the notion of elementary propagating degrees of freedom can change (e.g. some may become composites of more fundamental ones),  but at any scale there always exist some. 

 Resolving a distance scale $L$ means that we integrate in propagating degrees of freedom of mass/energy  $1/L$, which  can be treated as elementary at distances $L$.  For example,  it should make sense to talk about 
 interactions of these degrees of freedom within the space-time interval of size $L$. 
   All known non-gravitational UV-completions are based on this fundamental notion. By extending this concept to UV-completions of gravity beyond the Planck length $L \, \ll \, L_P$ (or  more general   $L \, \ll \, L_*$),  one would try to integrate in some trans-Planckian degrees 
 of freedom of mass $1/L$.    However,  as suggested in \cite{dvali-gomez}, in Einstein gravity trans-Planckian propagating degrees  of freedom cannot exist, instead any such degree of freedom becomes a classical state with smallest size $\sim \, 2L_P^2/L$, the Schwarzschild radius corresponding to the mass $1/L$.  This classical state  is no longer an independent entity and is fully described by already existing IR degrees of freedom, such as the massless graviton.  
 Thus, the would-be trans-Planckian states carry no information   about the trans-Planckian physics and decouple from quantum processes, just as classical objects should do.
  In this way, Einstein gravity self-completes itself in the deep UV by mapping would-be 
 trans-Planckian degrees of freedom to classical IR states \cite{dvali-gomez}. In particular,  this is the field theoretic manifestation of the fact that  in  Einstein gravity  the Planck  length is the shortest length-scale of nature and the underlying reason for the so-called Generalized Uncertainty Principle \cite{GUP1},\footnote{In \cite{Giddingsloc}, this  obstruction to probe short distances has been suggested to be related to a kind of locality bound, where below that scale the local quantum field theory no longer captures all dynamics.}. This notion also exists in string theory where it can be argued that the fundamental string length as well sets a limit on the shortest distance which is possible to probe  \cite{GUP2}.
 
The formation of BH as an outcome of Trans-Planckian collisions is a natural expectation (see e.g. \cite{thooft}). The discovery of low scale quantum gravity scenarios \cite{add,aadd} promoted this possibility to a potentially experimentally-observable phenomenon. Indeed, BH formation in high energy scatterings at particle colliders was predicted in \cite{aadd} (for subsequent work in this direction see \cite{others}). This feature of gravity was formulated in terms of the ``Asymptotic Darkness'' as a unique outcome of Trans-Planckian scattering at small impact parameter in \cite{Banks}.

In this work, we will furthermore claim that a BH is 
the only output of {\it any} trans-Planckian scattering          
process in {\it any} healthy theory of gravity. In other words we will argue that there is        
{\it no} contribution from sub-Planck distance physics in                           
{\it any} high (or low) energy scattering process.

 In the present paper we shall reiterate the above notion and, in this light, address the viability of attempts of conventional (Wilsonian) UV-completions of Einstein gravity in the trans-Planckian domain.
  
This understanding has important consequences for cases in which gravity is assumed to become weaker in the deep UV,   an example of which is the Asymptotic Safety Scenario \cite{Weinberg}. That is, the mapping of  trans-Planckian gravity  to classical IR gravity  excludes 
 UV-completions of gravity by asymptotically safe behavior in any domain. 
     
\begin{figure}[!t]
  \begin{center}
    \includegraphics[width=3.5in]{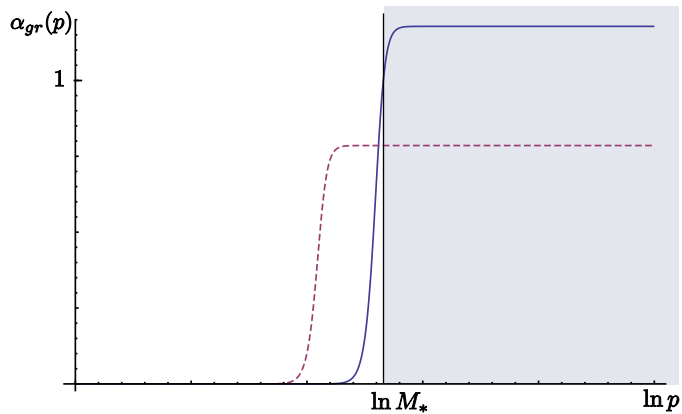}
  \end{center}
  \caption{ Momentum-scale dependence of $\alpha_{grav}$. The dashed line shows a running of the gravitational coupling where gravity becomes weaker in the weakly coupled regime. In a ghost free theory this cannot happen. The solid line represents a typical running of $\alpha_{grav}$ usually found within the Asymptotic Safety scenario. Here, gravity first hits the strong coupling ($\alpha_{grav}=1$) at scale $M_*$, before turning over to the fixed point scaling. The shaded region indicates the regime in which Black Hole formation takes place and which hence cannot be probed by experiments.}
  \label{running_gn}
\end{figure}

    In order to see this,  we will first show that gravity cannot become weaker than in pure Einstein gravity before  hitting the strong coupling scale, by requiring the absence of negative norm states. To be more precise, as shown in \cite{brustein-dvali-veneziano},   in any ghost-free theory in the weak-coupling domain $\alpha_{grav}(p)$ must satisfy,  
 \begin{equation}    
\alpha_{grav}(p) \, \geq \, \alpha_{Ein}(p) \, \equiv \, p^2/M_P^2 \, ,
\label{grow}
\end{equation}
and the quantity $\alpha_{grav}(p) / \alpha_{Ein} (p)$ must be  
a non-decreasing function of  $p^2$,  at least until  $\alpha_{grav}(p)$ becomes of order one. 
 In other words,   a weakening of gravity cannot set in while  $\alpha_{grav} (p)\, \ll \, 1$ (see Fig. \ref{running_gn}).  Thus,  before the turn-over, $\alpha_{grav}(p)$ first has to reach the strong coupling point.  Hence, the turn-over cannot happen 
for $p\, \ll \, M_*$. 

However,  beyond $M_*$ we are in the trans-Planckian domain, which is mapped on classical IR  
gravity.   So gravity cannot display Asymptotic Safety in any well-defined physical sense.  
Due to the Black Hole (BH) barrier, beyond $M_*$ any shorter distances cannot be probed in principle.  
 For instance,  in the  trans-Planckian domain scattering cross sections with center of mass  energy $E$ should be dominated by BH-production, which can be estimated to be given by the geometric cross section $\sigma\, \sim \, E^2/M_P^4$. This growth is hard to reconcile with the notion of Asymptotic Safety, or with a weakening of UV-gravity in general. This result agrees with a complementary proof of the impossibility of AS in a theory of gravity containing BHs \cite{entropy}. In \cite{entropy}, Shomer shows that any UV fixed point in which gravity becomes weaker, as in AS, is incompatible with the Bekenstein-Hawking entropy for BHs. The observation that the BH barrier prevents probing  the fixed point behavior of Asymptotic Safety has also been made in \cite{GiSchmiSoAn}. 
 
    To summarize,  the self-completeness of gravity raises the question, whether Wilsonian UV-completions of trans-Planckian gravity are viable or even physically motivated in the light of \cite{dvali-gomez}. 
 
\section{Non-existence of Sub-Planckian Distances in Einstein Gravity}
\label{sec:no subplanck}

In this section, we shall reiterate the point of \cite{dvali-gomez}.  We shall first discuss why trans-Planckian physics, in the sense of probing distances 
   $L \, \ll \, L_P$,  cannot exist in Einstein gravity.
   
  First of all let us mention that this statement is Lorentz invariant (as seen from an observer at spatial infinity) as distances (and energies) here refer to the distances (and energies) measured in the center of mass reference frame. In this frame, one may also use the seemingly non-relativistic relation that shorter distances are measured by higher energies, i.e. $E\sim 1/L$. Of course, a boost will accordingly change the $L$ and $E$ values but not their relation. However, a reader may be worried that in the highly nondynamical gravitational background due to the collision of the sources, the definition of length should include some notion of the local spacetime. In this case we will always refer as a ``length'' the instantaneous local invariant length measured by an ADM observer \cite{wald}. In this case, the four dimensional metric is split in $3+1$ as
 \be
 ds^2=-N^2dt^2+g_{ij}(dx^i-N^i dt)(dx^i-N^i dt)\ .
 \ee 
 As we will only be concerned with S-wave (spherical) scatterings, at a fixed time one may then choose coordinates such that they define the following (instantaneous) three-dimensional metric \cite{hoop2}
\be
 g_{ij}\Big|_{t={\rm const}}=\phi^4(r)\delta_{ij}\ ,
 \ee
 where $\delta_{ij}$ is the Kronecker delta. In this way the invariant length is given by 
 \be
 L(r_0)\equiv4\pi\int_0^{r_0} dr \phi^{2}(r)\ ,
 \ee
 where $r_0$ is the coordinate radius we would like to measure.
 
From now on we will assume without loss of generality that we will just talk about distances $L$. In light of these definitions, to be precise with our opening statement, we will prove that instantaneous distances shorter than the Planck length cannot be probed.

\subsection{Field theoretical hoop conjecture}

In quantum field theory any measurement that attempts to resolve the distance  $L$ has to excite, via a scattering experiment, 
 degrees of freedom of  energy $1/L$ within the box of size $L$.  The explicit construction of such a scattering experiment would involve at least two particles which are boosted in such a way that their (Lorentz-invariant) center of mass energy exceeds $1/L$ and that their impact parameter will be less than $L$.  For $ L\, \ll \, L_P$ such an attempt will lead to the formation of a  classical BH (see also \cite{GUP2,Banks,aadd,others}). Note, that by itself none of the involved boosted particles is a BH even when boosted to energies $\gg M_P$ as no graviton exchange is involved. Their correct description is given by the so-called Aichelburg-Sexl geometries \cite{Aichelburg}. The fact that the outcome of such an experiment will inevitably produce a BH can be regarded as a field-theoretic  interpretation of Thorne's hoop conjecture \cite{tho}, according to which a BH with horizon forms when, and only when, a mass $M$ gets compacted into a region whose circumference in every direction is less than the corresponding Schwarzschild horizon
($R_S(M)$) \footnote{This version of the conjecture is of course very vague as it implies the
existence of an omniscient observer who can define a global event horizon.
However, although this conjecture can be generalized by introducing a local
definitions of horizons, i.e. closed trapped surfaces (see \cite{Eardley:2002re} and references therein, 
\cite{senovilla}), we will only be interested in the point of view of
asymptotic observers in flat space (S-matrix) where the above formulation
of the conjecture is applicable.}. In terms of such a scattering experiment, this means that a BH will form anytime the transfer energy is localized (dynamically) within $R_S$.  Thus an attempt of resolving sub-Planckian distances will lead to the formation of a macroscopic BH of horizon size $2L_P^2/L$, which can only probe large distances. 
This observation leads to two important conclusions. The first one is that a elementary state with mass bigger than $M_p$ cannot exist because its Compton wavelength is below $R_S$. the second conclusion is that, by BH barrier, no sub-Planckian distances may ever be probed and therefore it is not a physical statement to talk about these distances. 

The previous discussion has been however based on a classical analysis so one might wonder whether quantum mechanical arguments could spoil it. It has been argued in \cite{Banks} that a scattering experiment of transfer energy $E\gg M_p$, with impact parameter $L\ll L_p$, may indeed produce elementary particles as an outcome with (quantum) probability $e^{-E ^2 L_p^2}$. The key observation we want to make here is that such small probability is due to a production of a virtual BH. This conclusion can be drawn by noticing that the factor $E^2 L_p^2\sim S$ where $S$ is the Bekenstein-Hawking BH entropy and therefore the suppression $e^{-S}$ represent a Boltzmann suppression. In other words, the produced particles are the result of a BH which formed during the collision and subsequently evaporated completely into elementary particles in a short time.
Because the Compton wavelength of the emitted elementary particle is larger than the Planck length, this implies that again, even in this rare case, no sub-Planckian distances may be probed.

\section{Einstein Gravity is the Weakest Gravity}\label{EGWG}

  We have seen in the previous section that because of the BH barrier,  sub-Planckian distances are 
  unphysical, and therefore the only sense in which we can talk about gravity at trans-Planckian energies is in terms of  IR gravity. 
  
   This fact eliminates the need of a UV-completion of Einstein's theory which would be due to an improved 
   behavior of the graviton propagator  for large $p$.  
  
  In this section we will prove, following the reasonings of \cite{brustein-dvali-veneziano}, that any modification of gravity that does not propagate ghost degrees of freedom in the weak regime, always produces a stronger gravitational attraction. In this sense, modifying the theory of gravity only leads to a BH production at even lower energies making the BH barrier more efficient.

   For a scattering process of particles with characteristic momentum
transfer $\sim p$ and a center of mass energy $E\sim p$, weak gravity is defined as 
the condition
\be
  \alpha_{grav}(p) \, \ll \, 1,
\label{weakcondition}
\ee
where $\alpha_{grav}(p)$ is given by (\ref{Ap}).
For example, in the 
pion-nucleon scattering at QCD scale energies, Einsteinian gravity is weak. 

In this regime, consider a {\it one-graviton} exchange process between two energy-momentum sources $T_{\mu\nu}$ and $\tau_{\mu\nu}$.
The amplitude of this process in momentum space is given by
\begin{equation}\label{eq:amplitude A}
 A(p)={T}^{\mu\nu}(p)\Delta_{\mu\nu,\alpha\beta}(p){\tau}^{\alpha\beta}(p),
\end{equation}
where ${T}_{\mu\nu}(p)$ and ${\tau}_{\alpha\beta}(p)$ are the Fourier-transforms of the sources, and 
$\Delta_{\mu\nu,\alpha\beta}(p)$ is the graviton propagator in momentum space.

In Einsteinian gravity, in which the gravitational force is mediated by a single massless spin-2 particle, the tensorial 
structure of $A(p)$ is uniquely fixed:
\begin{equation}\label{eq:massless A}
 A_{massless}(p)=G_N \frac{T_{\mu\nu}(p)\tau^{\mu\nu}(p)-\frac{1}{2}T_\mu^\mu(p)\tau_\nu^\nu(p)}{p^2} \, .
\end{equation}

Notice, if, in the UV (or IR), gravity deviates from the Einsteinian theory, the structure of $A(p)$ will be different, but 
still \emph{extremely} restrictive. This follows directly from the spectral representation of the graviton propagator for which the most general ghost-free structure is, 
\begin{eqnarray}
 \label{eq:spec rep}
&A(p)&=T^{\mu\nu}\Delta_{\mu\nu,\alpha\beta}\tau^{\alpha\beta}=\nonumber\\
&=&\!\!\!\!\frac{1}{M_P^2}\left(\frac{T_{\mu\nu}\tau^{\mu\nu}-\frac{1}{2}T_\mu^\mu\tau_\nu^\nu}{p^2}+\int_0^\infty\!\!\!\!\!ds\rho_2
(s)\frac{T_{\mu\nu}\tau^{\mu\nu}-\frac{1}{3}T_\mu^\mu\tau_\nu^\nu}{p^2+s}+\int_0^\infty \!\!\!\!\!ds\rho_0(s)
\frac{T_\mu^\mu\tau_\nu^\nu}{p^2+s}\right) ,
\end{eqnarray}
where we have seperated the contributions from the massless spin-2, the massive spin-2 and the spin-0 poles. It is 
crucial to note that the absence of ghosts demands $\rho_2(s)\geq0$ and $\rho_0(s)\geq 0$, $\forall s$. In order to understand the meaning of $\rho_2$ and $\rho_0$ let us consider the ADM decomposition \cite{Wald} of the metric. The graviton can be decomposed into a spin 2 field $h_{ij}$ (the spatial metric), a scalar $N$ (the lapse) and a vector $N^i$ (the shift) ($i,j,...$ are 3-dimensional indices (with a positive defined metric) and $\alpha,\beta,...$ are the 4-dimensional indices). In the transverse-traceless gauge (that can always be taken because of the linearized diffeomorphism group), the kinetic term of the spin 2 part looks like $(\partial_\alpha h_{ij})(\partial^\alpha h^{ij})$.
This kinetic term has {\it no} sign ambiguities, dependent on the choice of the 4-dimensional signatures, the sign in front of it determines whether $h_{ij}$ is a propagating ghost or not. This sign is encoded in $\rho_2$. Of course, different to GR, for example the trace of $h_{ij}$ (the scalar degree of freedom) can propagate and the sign of its kinetic term is determined by $\rho_0$.
 
 Also the tensorial structure is fixed by the requirement of the absence of ghosts. 

Then, we are lead to a powerful conclusion: The running of $\alpha_{grav}(p)$ (or equivalently $G_N(p)$) can be understood in terms of $\rho_2(s)$
and $\rho_0(s)$,  and the positivity requirement automatically excludes a weakening of gravity in the weakly-coupled regime \cite{gia-strong, brustein-dvali-veneziano}. 
Indeed, using the spectral decomposition (\ref{eq:spec rep}) we can represent  $\alpha_{grav}(p)$ in the following form,  
  \begin{equation}
 { \alpha_{grav}(p)  \over  \alpha_{Ein}(p)} \, = 1 \, + \, p^2  \, \int_0^{\infty}\!\!\! ds\,{\rho_2(s) \over p^2 + s} \, ,
  \label{generalalpha}
  \end{equation}
  where  $\alpha_{Ein}(p) \, \equiv\, p^2/ M_P^2 $ is the strength of pure-Einstein gravity  and relativistic sources are used.  Due to the positivity of   $\rho_2(s)$,   ${ \alpha_{grav}(p)  \over  \alpha_{Ein}(p)}$ is  a never decreasing function larger than one, 
 \begin{equation}
  { \alpha_{grav}(p)  \over  \alpha_{Ein}(p)} \, \geq\, 1\, ~~~{\rm and} ~~~
  {d \over dp^2} \,  \left ( { \alpha_{grav}(p)  \over  \alpha_{Ein}(p)} \right )\, \geq\, 0\, .
  \label{ratio}
  \end{equation} 
 
Thus, weak gravity can never become weaker! In other words, Einsteinian gravity is the weakest among all possible gravities that flow to Einstein with a given $G_N$ in the IR. A direct consequence of this fact  is that, in the weak gravity regime, any modification of Einsteinian gravity always produces (for a given mass) BHs of size $R_H\ge R_S$, where $R_S=2G_NM$ is the Schwarzschild horizon
\cite{brustein-dvali-veneziano}.
 
 The physical meaning of the above statement is clear.  Equation (\ref{eq:spec rep}) tells us that  the gravitational force mediated  by positive norm 
 particles is always attractive. Thus,  the weakest gravity at any scale is the one that is mediated by the 
 minimal number of messengers; this is Einstein's gravity mediated by a single 
 massless spin-2 graviton.  Furthermore, the positivity of $\rho_2(s)$ and $\rho_0(s)$ requires  the strong coupling scale $M_*$ of any UV modification of gravity to be lower than the strong coupling scale of pure Einsteinian gravity,
 \be
 M_*\le M_P \, .
 \label{strongcouplingscale}
 \ee
This fact is a direct consequence of (\ref{ratio})  (see also appendix \ref{B}). 

An example of a healthy modification of Einsteinian gravity with such a property is provided by Kaluza-Klein theories in which gravity
becomes high-dimensional above a  compactification scale $M_c \equiv 1/R_c$.  
For example, the $5$-dimensional case corresponds to a particular case of (\ref{eq:spec rep}) with 
  \begin{equation}
  \rho_2(s) \, = \, \sum_n \, \delta (s - (nM_c)^2) \, , ~~~ \rho_0(s) \, =  \delta(s) \, .
  \label{5d}
  \end{equation}  
Then, for energies $p\gg M_c$, (ignoring 
tensorial structures)  the one-graviton exchange amplitude takes the form
\begin{equation}\label{eq:prop in 5d}
A(p)\propto\frac{1}{M_*\sqrt{p^2}}, 
\end{equation}
where $M_*$ is the 5-dimensional Planck mass. We can recast this as the usual 4-dimensional propagator but with a 
momentum dependent Newton constant $G_N(p)$
\begin{equation}
 \label{eq:4d prop}
A(p)\propto G_N(p)\frac{1}{p^2},
\end{equation}
where $G_N(p)\equiv \frac{\sqrt{p^2}}{M_*^3}$. At $p=M_c$, the ``running"  Newtonian coupling 
must  match the four-dimensional Newton's constant. This matching gives the well known geometric relation between the four-dimensional and five-dimensional Planck scales,  
\begin{equation}
 \label{eq:matching of Planck scales}
M_*=(M_P^2M_c)^{\frac{1}{3}}\leq M_P .
\end{equation}
Here, gravity becomes strong at a scale $M_* \leq M_P$, due to the fact that compactification scale is  smaller than the 4 dimensional Planck scale, $M_c<M_P$.

It is very important to note that what we just said is independent of the precise change of the law of gravity. Hence, the only hope for gravity to become weaker in the UV is to first reach the strong 
coupling regime (the scale $M_*$) and only then turn around its strength. 
We shall now try to see if such a behavior can be physically meaningful.

\section{Trans-Planckian Gravity is IR Gravity} 

In the previous section we proved that gravity cannot become weaker than GR in the weak gravity regime. 
Could we allow $\alpha_{grav}/p^2$ to start decreasing in the strong gravity region,  $p \, \gg \, M_*$? 
We shall now argue that such a decrease is  unphysical, because the 
region  $p \,\gg \, M_*$  is protected  by the BH barrier.   Because of the impossibility of probing length scales $L \, \ll \, L_*$,  an asymptotic weakening of gravity  at such distances is physically meaningless, and moreover 
 unnecessary,  since gravity in this region is fully controlled by large distance classical dynamics. 
 The key point is that there are no perturbative states with masses
   above  $M_*$ that can be excited. This is due to the fact that the Compton wavelength of such a particle would be smaller than its BH horizon. Rather, the only well-defined meaning for any such state is that of a non-perturbative classical object, probing at best distances of order of its own BH horizon, 
which because of the positivity of the spectral representation is always above $L_*$.

One may however wonder whether this result can be implemented in a consistent quantum field theory. In other words one may wonder whether            
imprints of sub-Planck-distance physics cannot be avoided in infinitely precise low energy experiments. If this were true than the existence of large mass quantum states would be necessary, invalidating our results. 
                                                                                                                                                                       
In  the Wilsonian picture the information                                                                                                                                                      
  about a large energy physics ($E=M\gg M_P$) is carried by {\it propagating} quantum degrees of freedom of mass                                                                                                                                                    
  $\sim  M$. By integrating-out such particles in the effective low energy theory, we get a                                                                                                                                                    
  series of operators of the form                                                                                                                                                                                                        
  \begin{equation}                                                                                                                                                                                                                         
  g \, F_{\mu\nu} {1 \over M^2 \, + \, \Box}  F^{\mu\nu} \,                                                                                                                                                                                
    F_{\alpha\beta}{1 \over M^2 \, + \, \Box}  F^{\alpha\beta} \, + \, \ldots ,                                                                                                                                                              
    \label{photoneff}                                                                                                                                                                                                                      
    \end{equation}                                                                                                                                                                                                                         
where $g$ is some effective coupling of order one.                                                                                                                                                                                                      
 By performing very precise  measurements at  very low energies, we could in principle read-off                                                                                                                                            
 the structure of these operators and thus decode physics at distances $\sim M^{-1}$.  In our field theory, we however postulate that gravity is not Wilsonian and in fact that massive (trans-Planckian) states are classical BHs.
 
 In this picture, integrating out such classical objects, we get an operator that looks like the leading order (at low energies) of (\ref{photoneff}), or
  \begin{equation}                                                                                                                                                                                                                         
  g e^{-\frac{M}{M_p}} \, F_{\mu\nu} {1 \over M^2}  F^{\mu\nu} \,                                                                                                                                                                                
    F_{\alpha\beta}{1 \over M^2}  F^{\alpha\beta} \, + \, \ldots .  
    \label{photoneff2}                                                                                                                                                                                                                                                                                                                                                                               
    \end{equation}                                                                                                                                                                                                                         
The above form is due to the fact that we are considering virtual BHs into scattering of photons. Extrapolating from the properties of on-shell BHs we expect that also virtual BHs are thermodynamical objects and therefore their contribution into the scattering amplitude is at least Boltzmann suppressed. This produces the exponentially small factor $e^{-\frac{M}{M_p}}$. However, while the operators in (\ref{photoneff}) incorporate propagating degrees of freedom that show up at the next to leading order in the expansion $\frac{\Box}{M^2}$, in our case, objects of mass larger than $M_P$ are no propagating. Therefore, operators in (\ref{photoneff2}) differs greatly from the operators in (\ref{photoneff}). In fact, the operators in (\ref{photoneff2}) carry no more information about energy scale above $M_P$ than                                                    
 any other operator obtained by integrating-out any classical (or solitonic) object of mass $M$.                                                                                                                                         

\begin{figure}[!t]
\begin{minipage}[c]{0.49\textwidth}
  \begin{center}
    \includegraphics[width=2in]{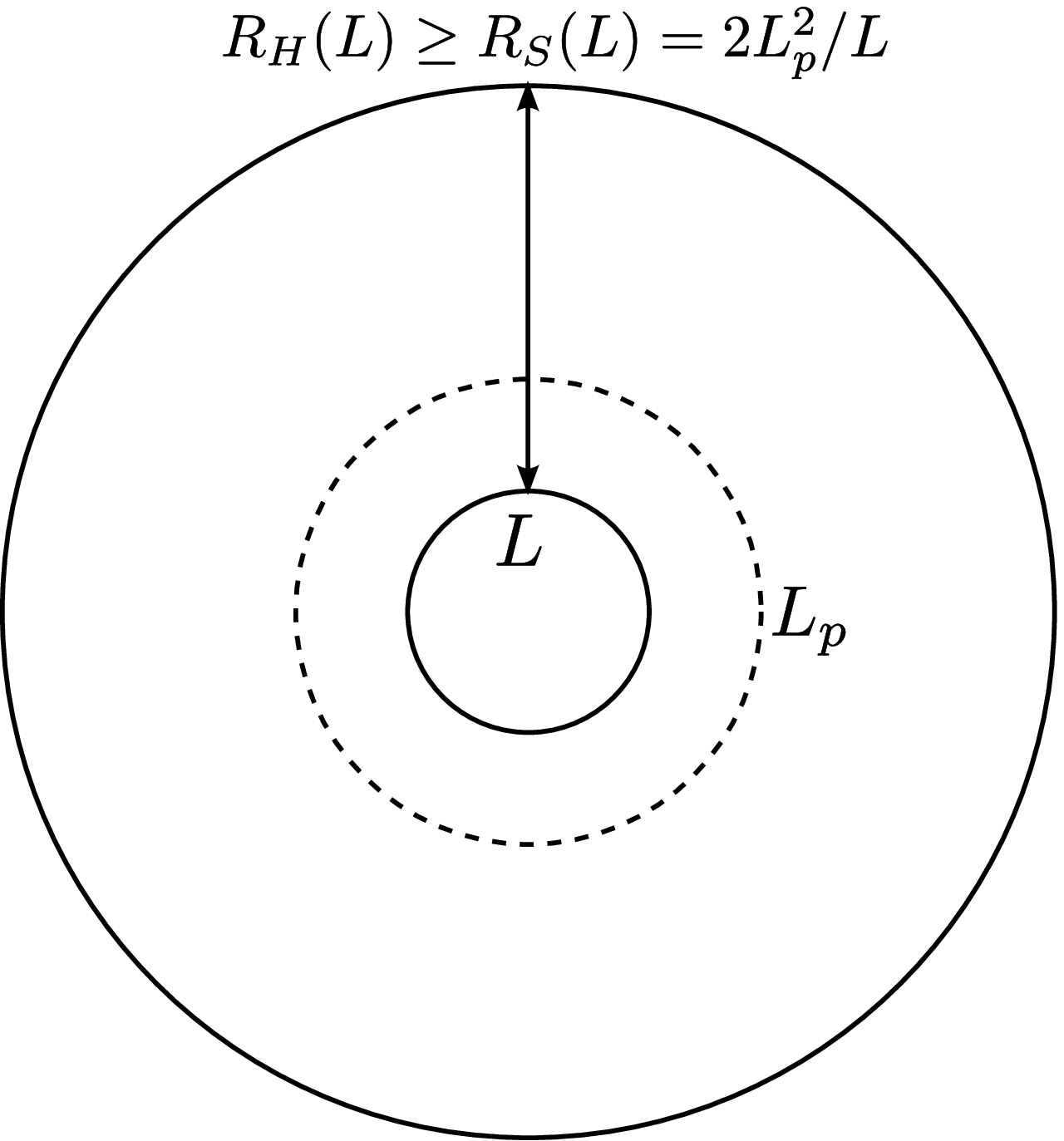}
  \end{center}
\end{minipage}
\begin{minipage}[c]{0.01\textwidth}
\end{minipage}
\begin{minipage}[c]{0.5\textwidth}
\begin{center}
\vspace{-1mm}\includegraphics[width=2.9in]{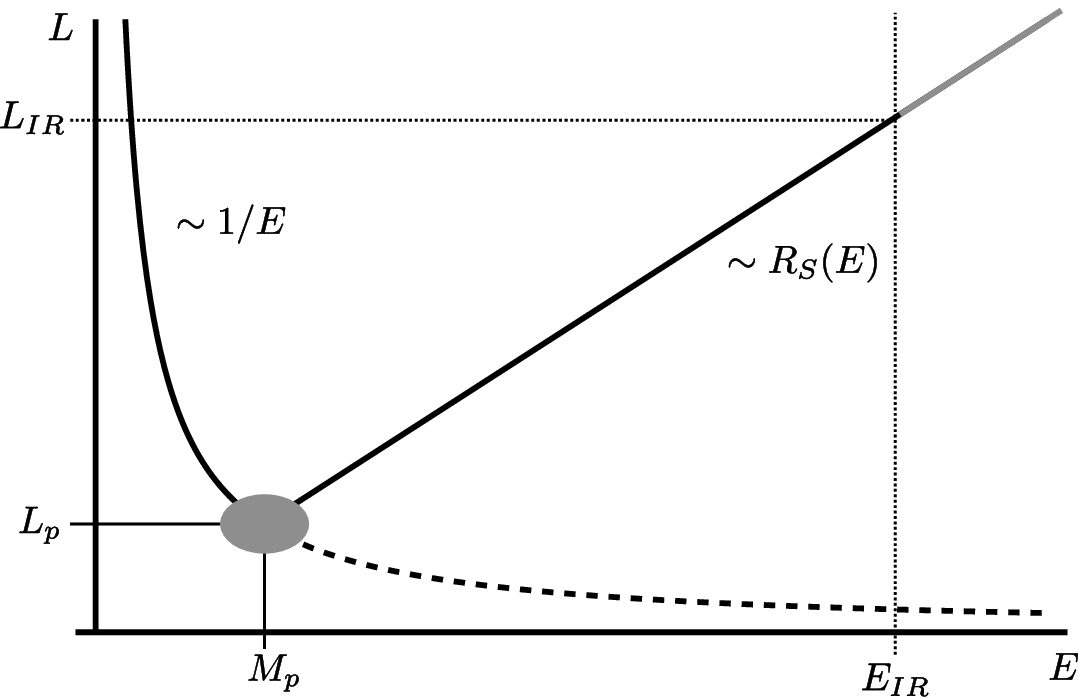}
  \end{center}
\end{minipage}
\caption{ Trans-Planckian distances are shielded by a Black Hole barrier. Probing poles at $p^2=L^{-2}\ll M_P^2$ one has to localize energy of order $L^{-1}$ within the distance $L$. The corresponding Black Hole horizon of this energy, $R_H(L)\geq R_S(L)=2L_P^2/L$, shields the sub-Planckian region ($L<L_P$) from being probed by any physical experiment. The sub-Planckian distance $L$ is mapped to the macroscopic distance $R_H(L)$. On the right-hand-side we show a qualitative plot of the energy-distance relation. The grey ``blob'' around the Planck scale indicates that at the Planck scale itself we don't know how the precise relation between engery and distances is. Also, there is an uncertainty about the far IR Black Holes, i.e. for energies $E_{IR}=2L_P^2/L_{IR}$ as we cannot exclude the possibility that at scales $L\gg L_{IR}$ gravity is modified.}
 \label{fig:schwarzschild}
\end{figure}

Summarizing, it is then clear that scales smaller than the strong coupling scale of the gravitational theory can never be resolved. This is in the same spirit of the Heisenberg uncertainty principle and in fact it is a high energies generalization of it (for a pictorial representation see Fig. \ref{fig:schwarzschild} and for previous literature see \cite{GUP1}). 

\subsection{Trans-Planckian pole in Einstein gravity}

 We want to discuss a concrete example in which one attempts to add an extra propagating degree of freedom to the  massless graviton in the trans-Planckian region. We show that if one tries to discover the trace of this new state in precise measurements at large distances, one inevitably fails. Let us try to modify the laws of  UV gravity by adding a scalar graviton $\phi$ of mass $m$ to the spin-2 Einstein graviton, $h_{\mu\nu}$.  That is, we shall assume that the metric seen by the external probe-sources is given by
 \begin{equation}
 g_{\mu\nu} \, = \, \eta_{\mu\nu} \, + \, {h_{\mu\nu} \over M_P}  \, + 
 \,  \eta_{\mu\nu}  {\phi \over M_P} \, .
 \label{metric}
 \end{equation}   
     We will see that this addition is meaningful only as long as $m \,  \lesssim \, M_P$
     and becomes meaningless when $m$ crosses over into the  trans-Planckian region.

\begin{figure}[!t]
  \begin{center}
    \includegraphics[width=3.4in]{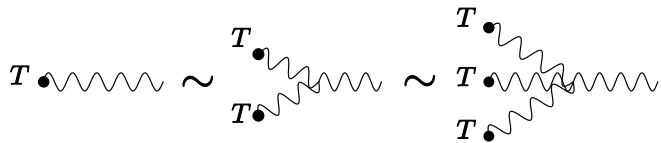}
  \end{center}
  \caption{Gravitational field produced by a source $T$. The wiggled lines represent the emitted gravitons $h_{\mu\nu}$. At the horizon the trilinear and higher order interactions are of the same order as the one-particle exchange}
  \label{fig:nonlinear}
\end{figure}

 At large distances,  the dynamics of the massless spin-2 graviton is described by Einstein's equation,
 \begin{equation}
    G_{\mu\nu} \, = \, 8 \pi G_N\, T_{\mu\nu} \, ,
    \label{einstein} 
 \end{equation}
  which to linear order in the graviton can be written as 
(we use harmonic gauge $\partial^{\mu} h_{\mu\nu} \, = \, {1 \over 2} \partial_{\nu} h$),   
  \begin{equation}
  \square  \, h_{\mu\nu} \, = \, - \, 16\pi \, G_N \, (T_{\mu\nu} \, - \, {1 \over 2} 
  \eta_{\mu\nu} \, T_{\alpha}^{\alpha} \, ),
   \label{gravitoneq}
  \end{equation}
  where $h\, \equiv \, h_{\mu}^{\mu}$.  To linear order  in the fields,  the only contribution to  
  $T_{\mu\nu}$ is coming from 
  the energy-momentum tensor of the external source, which will be taken to be a static, pointlike  mass $M$ with  
  $T_{\mu\nu} \,  = \, \delta_{\mu}^{0}\delta_{\nu}^{0} \, \delta^3(r) \, M$.  
  This gives the usual first order result for the graviton, 
  \begin{equation}
   {h^{(1)}_{\mu\nu} \over M_P}  \, = \, \delta_{\mu\nu} \, {R_S \over r} \, ,
  \label{lineargraviton}
    \end{equation}
where $R_S=2G_NM$ is the Schwarzschild horizon of the corresponding mass $M$ BH.  Note that to this order, the signal of approaching the horizon is that $h^{(1)}_{\mu\nu}$ becomes of order one. 
  At the same time, by consistency, the proximity of the horizon is signalled by the second and higher  order perturbations in $G_N$ becoming of the same order, i.e. the contributions from the non-linear coupling of the graviton to the source are becoming as important as the ones from the linear coupling to the source.  Hence,  the series has to be  resummed; see also Fig. \ref{fig:nonlinear}. This signals the formation of a horizon \cite{Duff}. 
   
   Notice, despite the corrections to the metric becoming of order one, the characteristic 
   momenta flowing through the graviton vertices are of order $1/R_S$, and thus,   
  as long as $R_S\, \gg \, L_P$, the near horizon geometry is not a probe of Planckian physics. 
  For such sources, gravity is in the weakly-coupled regime ($\alpha_{grav}\, \ll \, 1$).  
   
  In order to find non-linear corrections, we have to expand eq. (\ref{einstein}) to second order in 
  $h_{\mu\nu}$,  which effectively takes into account the interaction of the graviton with its own energy-momentum tensor $T_{\mu\nu}(h)$. To be fully consistent one also has to include the corrections to the energy-momentum tensor of the source $T^{(1)}_{\mu\nu}$ which is calculated in appendix \ref{A}. 
 \begin{eqnarray}
 8\pi G_N \,  T_{\mu\nu} (h) = && 
 - {1\over 2} \, h^{\alpha\beta} \left ( 
 \partial_{\mu}\partial_{\nu}  h_{\alpha\beta} \, + \, \partial_{\alpha}\partial_{\beta}  h_{\mu\nu}
 \, -  \partial_{\alpha}(\partial_{\nu}  h_{\mu\beta}  \, + \, \partial_{\mu}  h_{\nu\beta})
 \right ) \, - \nonumber \\
&&  - {1 \over 2}  \partial_{\alpha} h_{\beta\nu} \partial^{\alpha} h^{\beta}_{\mu} 
   +    {1 \over 2}  \partial_{\alpha} h_{\beta\nu} \partial^{\beta} h^{\alpha}_{\mu}  \,
   -\, {1 \over 4} \, \partial_{\mu} \, h_{\alpha\beta} \, \partial_{\nu} \, h^{\alpha\beta} 
 \nonumber \\
 &&    
   - \, {1 \over 4} \, \eta_{\mu\nu} \, \left ( 
   \partial_{\alpha} \, h_{\beta\gamma} \, \partial^{\beta} \, h^{\alpha\gamma}  \, - 
   \, {3 \over 2} \,  \partial_{\alpha} \, h_{\beta\gamma} \, \partial^{\alpha} \, h^{\beta\gamma} )
   \right )  
   \nonumber \\
   &&
        \, - \, {1\over 4} h_{\mu\nu} \, \square \, h \, + \,  
  {1\over 2} \,  \eta_{\mu\nu} \, h_{\alpha\beta} \, \square \,  h^{\alpha\beta}\,+ T^{(1)}_{\mu\nu},   
  \label{gravitonsource}
 \end{eqnarray}     
     evaluated for the linearized graviton $h^{(1)}_{\mu\nu}$. We then  
 get the standard corrections to the metric at second order in $G_N$. For example, 
  \begin{equation}
   {h_{00}^{(2)} \over M_P}  \, = \,  -{1 \over 2} \, {R_S^2 \over r^2} \qquad  \mathrm{and} \qquad {h_{00}^{(1)} \over M_P} = \frac{R_S}{r}(1+ a \frac{M^2}{M_P^2})\, ,
    \label{bilineargraviton}
    \end{equation}
   where $a$ is a factor of order 1 and $M\ll M_P$.
   Taking into account the backreaction of the gravitational field the source produces on itself gives a small shift in the ``effective'' gravitational mass of the source particle, which can be safely neglected.
   These corrections are the manifestation of the fact, that at the horizon, $r = R_S$, the expansion of the metric in powers of $(R_S/r)$ breaks down, and the series has to be resummed.   
   
   The novelty due to the presence of the massive scalar graviton $\phi$, also coupled to the same static external source $T$,    is  that  to second order in $G_N$, $h^{(2)}_{\mu\nu}$ gets corrections also from the coupling to 
 the energy momentum tensor of $\phi$, 
 \begin{equation}
 T_{\mu\nu}(\phi) \, = \, \partial_{\mu}\phi \partial_{\nu}\phi \, - \, {1 \over 2}\, 
 \eta_{\mu\nu}
 (\partial_{\alpha}\phi \partial^{\alpha}\phi\, + \, m^2 \phi^2) \, .
 \label{scalarenergy}
 \end{equation}
 These corrections are accounted for by including  the contributions from (\ref{scalarenergy}) evaluated to the first oder solution  $\phi^{(1)}\, = \, e^{-mr} (R_S/r)$ on the right hand side of  (\ref{einstein}).  Obviously,  this contribution gives only  an exponentially-supressed correction  
 to $h^{(2)}_{\mu\nu}$. 
 
  The more important,  power-law-suppressed, corrections can appear  if there are couplings between $\phi$ and $h$ of the form,  
 \begin{equation}
   { \phi  \partial^n h^k  \over M_P^{n+k-3}} \, , 
   \label{operators}
   \end{equation}
 as in non-minimally coupled gravity and where the tensorial structure is not disclosed. 
   Such couplings will induce an effective source for $\phi$,
   \begin{equation}
  ( \square \, + m^2 \, )\, \phi \, = \,  {  (\partial^n h^k)  \over M_P^{n+k-3}} \, + \, ... \,  , 
   \label{phisource}
   \end{equation}
  and can give corrections to $\phi$  which are not exponentially suppressed, but only by powers of  $(rm)^{-1}$ and $(rM_P)^{-1}$.   For example,  evaluating the right hand side of  (\ref{phisource}) for 
  $h \, = \, h^{(1)}$ and $r\gg m^{-1}$ can (subject to cancellations in the tensorial structure) give corrections 
  of the order 
  \begin{equation}
 {\phi^{(k)} \over M_P} \, \sim \,  {R_S^k \over r^k} {1 \over (M_Pr)^{n-2} (mr)^2} \, .
 \label{powercorrections}
 \end{equation}   
  The reason why these correction are not exponentially suppressed can be understood from the fact  that they arise  from short range processes which do not require propagation  of virtual $\phi$-quanta over distances larger than their Compton wave-lengths.
 
\begin{figure}[!t]
  \begin{center}
    \includegraphics[width=2.9in]{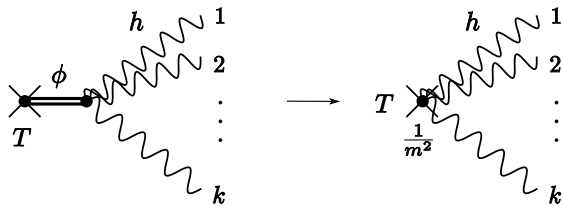}
  \end{center}
  \caption{A heavy scalar (double line) is mediating the interaction between a source $T$ and $k$ gravitons (wiggled line). Integrating-out this scalar at tree-level will induce an effectice point-like interaction between the source and $k$ gravitons. }
  \label{fig:integratingout}
\end{figure}

 In other words, these corrections can be viewed as the corrections to the metric
 in form of non-linear powers of exclusively massless gravitons, 
 appearing as a result of a tree-level integrating-out of a heavy scalar graviton of mass $m$, see also Fig. \ref{fig:integratingout},
  \begin{equation}
 g_{\mu\nu} \, = \, \eta_{\mu\nu} \, + \, {h_{\mu\nu} \over M_P} \,  + \eta_{\mu\nu}  { (\partial^n h^k)  \over M_P^{n+k-3}m^2} \, + \,  ... \ .
   \label{metriccorrected}
   \end{equation}
   
  To summarize, we have seen that corrections coming from a heavy gravitational degree of freedom to the Einsteinian metric at distances larger than its Compton wavelengths are 
  suppressed  either exponentially, or by inverse powers of its mass $m$ and cannot
  significantly affect Einsteinian gravitational dynamics at distances $r \, \gg \, m^{-1}$. 
  For example, they cannot interfere with the formation of  BHs with Schwarzschild radius 
  $R_S\, \gg \, m^{-1}$.   This is in full accordance with the notion of a decoupling of heavy states at low energies \footnote{We rely on the decoupling theorem, i.e., the theory at low energies does not give us any information on the theory at high energies.}. Although a heavy quantum state gives negligible corrections to the metric at large distances, in the case $m \, \lesssim \, M_P$, these corrections are measurable. In this way signatures of new gravitational physics of distance-scale $m^{-1}$  can in principle be probed by precision measurements at much larger scales $r \,  \gg \, m^{-1}$. 
  
  However, for $m \, \gg \, M_P$ this is not true as the new degree of freedom is no longer a perturbative state, but rather a macroscopic BH, which does not carry any information about the UV physics. Take for example the above massive scalar graviton $\phi$ of mass $m$.  Once $m \, \gg \, M_P$,   $\phi$ can no longer be treated perturbatively as its Compton wavelength ($m^{-1}$) is smaller than its BH horizon.  To see this, it is enough to examine the gravitational field produced by the non-relativistic particle $\phi$ by simply replacing $M$ by the mass $m$ in equation (\ref{einstein}). The analysis following eq. (\ref{einstein}) immediately shows us that the $\phi$ particle develops a horizon ($R_{\phi} \, = \, 2G_Nm$) at scales larger than $m^{-1}$. For this reason, it is a fully legitimate classical BH. 
Hence,  our perturbative analysis, in which we  considered contributions of virtual $\phi$ quanta, is no longer applicable. 
This also implies that the effective operator obtained by integrating-out $\phi$ can no longer have the power-law suppressed form. Instead, we must take into account that $\phi$ is now a BH, and therefore  any contact interaction resulting from its integrating-out must be exponentially suppressed by at least the entropy ($S$) factor $e^{-S}$. In other words, by becoming trans-Planckian, $\phi$ cannot carry information other than what is already carried by an IR black hole of the same mass.  Therefore, a particle with trans-Planckian mass should be integrated out as an ordinary classical BH of the same mass.

To summarize, given the fact that any degree of freedom with mass $M\gg M_P$ is a classical object, it becomes obvious that -- no matter how sophisticated one tries to be -- there is {\it no} process (including BH evaporation, primordial quantum fluctuation, scattering experiments, etc.) which can probe trans-Planckian physics.  

Note that the covering of sub-Planckian distances is even more efficient than what we described before. In fact, as shown in \cite{Gidd}, before the BH formation an eikonal barrier may well form. In this case eikonal  amplitudes (exchange of many soft gravitons) become important preventing hard energy transfers through a single graviton line which would encode information about short distance physics.

   \subsection{Trans-Planckian poles in general theories of gravity}
   
 In this section we extend the result of the previous section to the case in which the strong gravity scale is $M_*<M_p$.
  We will prove that, even in this case, no elementary states with mass  bigger than $M_*$ may exist. In order to prove that we will consider the one-graviton exchange analysis as a good approximation up to the strong coupling scale. In fact, although large classical BH are formed by a large number of gravitons (collective effect) as to produce a strong gravitational field, they can be treated separately as weakly coupled. 
          
     The proof is a direct consequence of the fact that the scale of the strong coupling  $M_*$ defined by 
 (\ref{strongscale})  also sets the upper bound on the center of mass energy  and the inverse impact parameter above which the BH formation starts. 
 
   The most straightforward way to prove this is to start from the opposite end.   Let us take a large classical BH of mass $M$ and horizon $R_H$.  The only condition on $R_H$ is that for momenta $p\, = \, R_H^{-1}$ gravity is in the weakly-coupled regime,  $\alpha_{grav}(p=R_H^{-1}) \, \ll \, 1$.   The relation between the horizon and $M$ can be found from the condition that 
 $h_{00}(R_H)/M_P \, = \, 1$, which using  (\ref{eq:spec rep})  can be rewritten as,  
 \begin{equation}
 \label{horizonnew}
\frac{h_{00}(R_H)}{M_P}\, = \, 2\int_0^\infty\!\!\!\!\!ds\frac{\rho(s)}{M_P^2}\frac{e^{-\sqrt{s}R_H}}{R_H}M\, = \, 1.
\end{equation}
Using the same (\ref{eq:spec rep}), we can also represent $\alpha_{grav}(p)$ as, 
\begin{equation}
  \alpha_{grav}(p) \, =  \, {(p^2)^2 \over M_P^2}  \, \int_0^{\infty}\!\!\!\!ds {\rho (s) \over p^2 + s} \,  .
  \label{generalalpha1}
  \end{equation}
We can now start decreasing the mass of the BH,  until the horizon and the inverse mass cross, $R_H \, = \, M^{-1}$.  We shall denote the corresponding mass by $M_* \equiv \, L_*^{-1}$.    Black holes heavier than 
$M_*$ are in the classical regime.   The  scattering processes with  center of mass energy
$ M\, \gg \, M_*$ and impact parameter $ \, \ll \, R_H$ will lead to classical BH formation.  
The crucial point is  that  the strong coupling is reached precisely around this energy $M_*$ and never well below.  In other words, there is no window above $M_*$ in which one can  probe $\alpha_{grav}(p)$  without encountering BH formation.  

   We can see this easily from the fact  that $\alpha_{grav}$ evaluated for momenta $p = M_*$ is of the same order as the quantity $h_{00}(M_*^{-1})/M_P$, 
\be
  \alpha_{grav}(M_*) \, =  \, {M_*^2 \over M_P^2} \, \int_0^{\infty}\!\!\!\!ds {\rho (s) \over 1 + s/M_*^2} \,  
  \sim 
\, \frac{h_{00}(L_*)}{M_P}\, = \, 2{M_*^2 \over M_P^2} \, \int_0^\infty\!\!\!\!\!ds\, \rho(s) \, e^{-\sqrt{s}/M_*}\, = \, 1 .
  \label{alphahorizon}
  \ee
This approximate equality follows from the fact that $\rho(s)$ is a positive definite function which gets exponentially cut off by the above-discussed Boltzmann factor e$^{-\sqrt{s}R_H}$, where  $R_{H}$ is the horizon 
of a classical BH of mass $M \, = \, \sqrt{s}$, determined from (\ref{horizonnew}).  So, effectively the integration is cut off at  $s \, = \, M^2_*$ (recall that for $M_*$, $R_H(M_*)=M_*^{-1}$), which makes the difference 
between the factors  $e^{-\sqrt{s}/M_*}$ and $(1 + s/M_*^2)^{-1}$ irrelevant.
 
 \subsection{On the weakening of gravity at strong-coupling scale}

By summing up the previous findings we are lead to the following picture: 
 By gradually increasing  $p$ in the scattering experiment we probe stronger and stronger 
 gravity.   By the time we reach the scale $M_*$, where gravity becomes strongly coupled, 
 we start seeing the BH formation. Any further attempt  of increasing $p$ will result in the formation of larger and larger classical BHs.  The region beyond $M_*$ is thus outside of physical reach.  
  Any  weakening of  $\alpha_{grav}(p)$ for  $p \, \gg \, M_*$  has no clear physical meaning since it cannot be probed.   

   Since we are only deriving the upper bound 
      on the threshold scale of BH formation, being $M_*$,  our proof is insensitive to the details of the theory.   We are approaching it from the weakly coupled 
 domain, in which the one-particle exchange is a good approximation, and stop 
 as soon as this approximation breaks down.   In this way, we manage to derive 
a necessary connection between the strong-coupling and  the threshold of BH formation, 
 which allows us to see the  impossibility of  probing physics at distances shorter than 
 $L_*$.     

 We shall now illustrate our general conclusion on two examples. 

\paragraph{An attempt of asymptotically safe gravity in four-dimensions} 

Consider a theory where Einstein gravity is valid all the way up to the Planck scale. 
 In such a theory $M_*\equiv M_P$. We then wish to modify the theory in such a way that
 in the deep UV ($p\rightarrow \infty$) a fixed point scaling for the gravitational coupling sets in, i.e. $\alpha_{grav}\rightarrow \alpha_\infty =const$. Let us see whether such a behavior could have a well-defined physical meaning. 
   This behavior can be modelled by a propagator of the form
\begin{equation}
 \label{eq:4d fixed pt prop}
\Delta(p)=\frac{1}{M_P^2p^2}\frac{1}{1+\frac{p^2}{\alpha_\infty M_P^2}}.
\end{equation}
where $\alpha_\infty> 1$ is a constant and $\Delta(p)\rightarrow \frac{\alpha_\infty}{p^4}$  for $p\gg \sqrt{\alpha_\infty}M_P$. In this limit $\alpha_\mathrm{grav}(p)=
16\pi G_N(p)p^2\simeq \alpha_\infty> 1$. This simulates, for example, the running of $\alpha_\mathrm{grav}$ of the Asymptotic Safety scenario. For probing distances $r\sim \frac{1}{\sqrt{\alpha_\infty}M_P}$ we need a center of mass
 energy of order $E\sim \sqrt{\alpha_\infty}M_P$ and a momentum transfer $p\sim E$. 

This example is very similar to the one of an additional graviton of trans-Planckian mass 
$m$, which we have considered previously, with the only difference that now the trans-Planckian state has a negative norm.  We shall ignore this sign for a moment since,  
despite this difference, our argument about the impossibility 
of resolving the heavy pole remains unchanged.  Therefore, any attempt of probing the length-scale  $L \, = \, \sqrt{\alpha_\infty}^{-1} L_P$ (corresponding to the  asymptotically safe 
 regime) will result in the formation of 
 a BH of macroscopic size, $R_H \, \simeq \, 2L_P\sqrt{\alpha_\infty}$. This BH formation cannot 
 be influenced by the  would-be asymptotically safe behavior in the deep UV, since    
 for the dynamics of a BH of size $R_H$, corresponding to $E\sim\sqrt{\alpha_\infty}M_P$, the ghost pole is decoupled and hence irrelevant. 
This is also obvious from the equality that determines the BH horizon,   
\begin{equation}
 \label{eq:Rs for example a}
h_{00}(R_H)=2\frac{\sqrt{\alpha_\infty}}{M_P}\frac{1}{R_H}\left[1-\alpha_\infty e^{-\sqrt{\alpha_\infty}M_PR_H}\right]=1 \, ,
\end{equation}
where we see that the existence of the heavy ghost pole at $\sqrt{\alpha_\infty}M_P$ only affects the value of $R_H$ with exponentially weak corrections. 
So in the attempted scattering process, a BH will be produced with radius $R_H\simeq
2\frac{\sqrt{\alpha_\infty}}{M_P}> M_P^{-1}$,  
which makes a penetration of the trans-Planckian region impossible. 
Thus Asymptotic Safety is rendered irrelevant before it had any chance to influence gravitational physics. 

To conclude, the existence of a ghost pole, which was assumed to be responsible for the would-be Asymptotic Safety behavior, is at best unphysical. 
  Moreover,  the UV-IR connection of gravity indicates that it should not have been included in the first place.   Indeed, because of the BH barrier, any physically sensible trans-Planckian state is mapped to a macroscopic object from the IR region.   However, in a consistent theory of gravity there are no negative energy classical states.  Therefore, the ghost pole simply cannot have any IR counterpart, and thus should be excluded by self-consistency  of the theory.

\paragraph{Asymptotically safe gravity with a lower cut-off scale}  

Next, we wish to consider an extension of the previous example in which gravity becomes strong 
at scale $M_*<M_P$. This will happen whenever new (positive norm) gravitons open up at some intermediate energies. 
 A good example of this property is provided by KK theories,  in which gravity becomes higher-dimensional 
 above the compactification scale $ M_c \, = \, R_c^{-1}$. Thus, at short distances $r \, < \, R_c$, gravity crosses over to the five-dimensional  regime and becomes strong at distances of the five-dimensional Planck length $L_*$. 
 
Let us then consider the case in which four dimensional gravity becomes weaker at high energies (a behavior shared with asymptotically safe gravity), i.e. at distances $\ll\sqrt{\alpha_\infty}^{-1}L_P$, and which has a strong coupling scale $M_*<M_P$.  To simulate this behavior we consider the following propagator
\begin{equation}
 \label{eq:5d mod prop}
\Delta(p)=\left\{\sum_{n=1}^{(M_P/M_*)^2}\frac{1}{p^2+\frac{n^2}{R_c^2}}\right\}\frac{1}{1+\frac{p^2}{\alpha_\infty M_P^2}},
\end{equation}
where $R_c M_*^3\equiv M_P^2$. The only difference to the previous example is that above the scale $\frac{1}{R_c}$ there 
is a tower of massive gravitons, which makes gravity strong at the scale $M_*$ as opposed to 
$M_P$.   The shortest observable length scale in this theory is  $L_*\equiv M_*^{-1}$.  There is again a trans-Planckian ghost pole, which makes gravity asymptotically safe, but because of BH formation, this pole is unphysical and cannot be probed.  
Indeed for energies required in order to probe the ghost pole,   $E\sim \sqrt{\alpha_\infty}M_P$,  the BH horizon is macroscopic $R_H\gg M_*^{-1}$, and the corresponding states belong to the classical gravity region.

  Again, we see that Asymptotic Safety has no physical meaning in this example.  The BH barrier, which  maps the trans-Planckian region to classical IR gravity, completely washes out the Asymptotic Safety region.

\subsection{Continuum Tails of Trans-Planckian Physics}

Previously,  arguing along the lines of \cite{dvali-gomez}, we have shown that trans-Planckian states cannot be detected in precision measurements at large distances.  The reason is that, because of the BH barrier, trans-Planckian states themselves are macroscopic 
 objects and are fully described by the classical IR sector of the theory. 
  Thus, their influence  must be fully accounted for by the large-distance gravitational physics. 
   We have illustrated this on the examples of isolated poles in the graviton propagator. 
  
  We wish to  extend this discussion to the continuum of states.    
Such states could result in sub-leading corrections to the one particle exchange diagrams represented by the decomposition (\ref{eq:spec rep}), which seemingly may be probed in the deep 
IR.

In particular, let us focus on sub-leading corrections that would make gravity slightly weaker. 

Without loss of generality, let us consider GR with a strong gravity scale $M_P$ and a perturbation to the Newtonian potential of the type
\be
\label{potential}
V(r)=G_N\frac{m M}{r}(1-\frac{L^2}{r^2}+{\cal O}{\left(\frac{L^3}{r^3}\right)})\ .
\ee

In fact, this potential has been considered for example as the correction to the Newtonian physics within the Asymptotic Safety scenario for gravity \cite{bh}. The 1-loop correction to the Schwarzschild metric in an effective field theory approach studied in \cite{Donoghue} was also found to be of this type.

  The negative  contribution $\propto \, L^2/r^3$  can be understood as the result of an exchange  of a continuum tower of 
 ghosts states.   
To see  this  we  can rewrite (\ref{potential}) as \cite{rs}, 
\be
\label{eq:horizon}
\frac{V(r)}{m}\simeq G_N M\left[\frac{1}{r}-L^2\int_0^\infty d\tilde{m}\frac{e^{-\tilde{m}r}}{r}\tilde{m}\right]\ .
\ee
The second term in the square brackets is nothing else than a sum over a continuum of massive particles. Indeed in Fourier space one readily obtains
\be
\frac{V(p)}{m}\simeq G_N M\left[\frac{1}{p^2}\, - \, {L^2 \over 2} \, \int_0^\infty ds\frac{1}{p^2+s}\right]\ .
\ee
In other words, the potential (\ref{potential}) is obtained by the exchange of a massless graviton and an infinite tower of equally distributed massive ghosts, i.e. with a constant spectral density $\rho(s)=L^2/2$. This potential in momentum space can also be considered as being a consequence of the following running Newtonian constant
\be
G_N(p)=G_N[1-L^2p^2\ln p]\ ,
\ee
which would make gravity weaker and weaker at high energies. 

 If  $L \, \gg  L_P$, the  contribution from the ghost tower  dominates  at the scale  $ \gg L_P$, and the theory makes no sense already in IR. So such a continuum correction cannot exist in a consistent theory.
 
  If  $L \, \ll \, L_P$,   for any classical BH the correction from the ghost tower  is subdominant to the 
  first non-linear correction from the massless graviton, which according to (\ref{bilineargraviton})  goes as 
  $\sim \, R_s^2/r^2$. Thus the ghost tower cannot affect the BH formation and the BH barrier cannot  be altered.  
  
   Consequently, the  trans-Planckian members of the continuum are again shielded by the BH barrier
   and are either unphysical or simply inconsistent. 
   
    Hence, there is no domain in which the potential (\ref{potential}) is a sensible description of physics.

A seeming way out of the impossibility of probing UV distances would be to construct a scattering experiment 
with center of mass energy $E<M_P^2 L$ and impact parameter $b<L$. In this region, it seems from the expanded potential (\ref{potential}) that BHs cannot form. However, such an experiment cannot be performed in principle. In fact, the impact parameter would be smaller than the Compton wavelength of a particle of mass $M=E$ and therefore, by Heisenberg principle, these center of mass energies $E$ can never probe distances $b$. The minimal distance that such an experiment can probe is $E^{-1}>(L_P/L)L_P\gg L_P$. So, once again, the Planck scale turns out to be impenetrable. 

\subsection{Sub-Planckian Experiments}                                                                                                                                                                  
In this section we would like to consider the question of whether trans-Planckian physics                                                                                                                                                                    
can be observed by                                                                                                                                                                    
preparing a scattering experiment at scales in which the modification of gravity is                                                                                                                                                                                 
already important.                                                                                                                                                                                                                             
                                                                                                                                                                                                                                           
In the weak gravity regime, as proven before, any healthy modification of                                                                                                                                                                      
gravity can only make gravity stronger                                                                                                                                                                                                     
and therefore the BH barrier would be even more effective than in the GR                                                                                                                                                                   
case.                                                                                                                                                                                                                                      
                                                                                                                                                                                                                                           
One may, however, envision an experiment prepared at distances shorter than the strong gravity                                                                                                                                                                      
scale where one can wonder whether the hoop conjecture may be violated and enable a resolution of the                                                                                                                                                                           
sub-Planckian scales without the BH barrier.                                                                                                                                                                 
Our results disagree with this point (see also \cite{later} and note \footnote{Although we agree with the result of \cite{later}, we disagree with the authors proof. In \cite{later} the effects of sub-Planckian distance physics has been simulated by an energy varying Newtonian constant $G_N(E)$. In this case, the gravitational mass would involve a volume integral of $G_N(E)$ producing a large effect which was ignored in \cite{later}.}). In fact, by conservation of energy, any                                                                                                                                                                    
experiment prepared at distance-scales smaller than the Planck length (and correspondingly energies                                                                                                                                                         
bigger than the Planck energies) has to be surrounded by a classical BH.                                                                                                                                                                              
                                                                                                                                                                                                                                           
Suppose such an experiment is indeed set up without a BH formation so that an asymptotic observer detects some output from the experiment.                                                                                                                                                                  
Then, this implies that a degree of freedom with energy larger than the                                                                                                                                                                    
Planck mass has to cross outside a sphere of radius $L_P$ before reaching the detector.                                                                                                                                                             
As soon as this happens, a BH will inevitably form, as discussed before. However, if there was no BH before                                                                                                                                                  
this would be in contradiction with the Êconservation of energy which is manifest on an asymptotically flat background.                                                                                                                                       
An asymptotic observer may in fact draw a sphere around the region of the experiment and continuously monitor the energy inside the sphere by measuring the Gaussian flux at infinity. Therefore, the only way of conserving the flux at infinity is to exclude the absence of the BH during any exchange of information from inside the sphere.
                                                                                                          
We thus conclude that {\it at best} the experiment was prepared inside of an already existing BH.                                                                                                                                              
                                                                                                                                                                                                                                           
From a slightly different perspective: One cannot have an energy                                                                                                                                                                   
localized within distance $L\ll L_P$ and crossing outside this region without causing a surrounding gravitational field of a BH.
                                                                                                                                    
Imagine for example an extreme case in which the gravitational force vanishes at some scale $L\ll L_P$.                                                                           
Take a spherical shell placed entirely inside this region.                                                                                                                                                                    
Although this shell has a positive energy it does not gravitate as long as its radius $R\ll L$. Naively then, one would conclude that there is no gravitational field outside the sphere, thus an asymptotic observer would see a flat space.  Now let the shell communicate with an outside observer by expanding and crossing outside the $L$ sphere. At some point this sphere will cross into a $R> L$ region. Since the energy of the shell is trans-Planckian it has to form a BH. However, the BH cannot appear out of nothing by conservation of Gaussian flux at infinity.
We then see that a BH must have been formed from the very beginning when the experiment is set up.                                                                                                             
                                                    
\subsection{Infrared Scales}

   For definiteness,  our treatment was limited to theories which  flow to Einstein gravity on 
 asymptotically-flat spaces  in the deep IR.    A theory may contain an infrared scale $L_{IR}$ beyond 
 which this assumption breaks down.  For example,  this may be a scale of a small background curvature,
 or something more profound.  
    Since the existence of such a scale may modify the properties of BHs,  the  connection between 
    deep-UV gravity and classical IR BHs can also be affected. 
     In such a case we still expect our conclusions to hold true in the energy interval 
   between   $1/L_{IR}$  and   $L_{IR} /(2 L_P^2)$, the latter value being set  by the  mass of 
   an Einsteinian  BH with Schwarzschild radius equal to $L_{IR}$ (see also Fig. \ref{fig:schwarzschild}). 
   
    If  $L_{IR}$  is a curvature radius  produced by a positive cosmological constant, 
   we expect the concept of a minimal length to be unaffected.  If anything, positive 
    curvature makes it harder to probe short distances, since for a given mass the effective  Schwarzschild radius is increased.  For example 
  the time component of a static metric would then be (in Schwarzschild coordinates)
\be
g_{00}=1-2G_N \frac{M}{r}-\frac{ r^2}{L_{IR}}\,.
\label{L}
\ee
For the observed cosmological constant, $L_{IR} \, = \, 10^{28}$cm,  the deviation from the flat space case only appears for energies comparable to the mass of the observable Universe and can be ignored.  
 
 If, for instance, $L_{IR}$ is related to a negative cosmological constant, one should consider the concept of  AdS/CFT correspondence \cite{ADS}. However, this will not be the concern of our paper.

\section{Conclusions}

In \cite{dvali-gomez}, it has been argued that quantum gravity might be fully described by light degrees of freedom. In this sense GR has been considered self-complete.
Following this idea we have shown that this self-completeness property, which is built-in in Einstein gravity, persist for a wide class of its deformations.  The basic reason for self-completeness is the non-existence of trans-Planckian  propagating degrees of freedom.  Any
      would-be trans-Planckian pole is mapped on a classical IR state, described by
  low-energy degrees of freedom of the IR theory.  This remains true in the presence of a BH if one assumes that BH evaporation is fully described by low energy physics.
  This assumption is backed up by noticing that, in order to see trans-Planckian corrections to the Hawking evaporation, one should integrate-in an operator of mass larger than the Planck scales. However, this operator defines a particle of Compton wavelength smaller than the BH horizon of the same mass in the weakly coupled region, so such an operator can only integrate-in other BHs.

As a consequence we have seen, the same properties that
 make Einsteinian gravity self-complete in the deep UV, also render  many attempts of a conventional
 UV-completion in the trans-Planckian region physically meaningless.

    We have focussed on the class of such attempted UV-completions which are based on the ideas of an asymptotic weakening and of Asymptotic Safety.    We have shown that in Einstein gravity and its ghost-free deformations there is essentially no energy interval in which these ideas can be realized in a physically clear way.  We have found  that, in both cases, the necessary condition is that  a weakening (or safety) can only take place within the strong gravity domain.
 We have shown that in ghost-free extensions of Einstein  this domain includes
 the domain where gravity starts to be mapped to the IR region, because of
the BH barrier.  In other words, there is no interval of distances in which gravity
may be strongly-coupled but not shielded by the BH barrier.

 Thus, both mechanisms of the completion are necessarily pushed in the region in which gravity seems to be  self-complete anyway.   So the only meaning of such completions would be if they are mapped to IR physics, but this is not possible for the asymptotically weak gravity case.

In this paper  we did not  address the question of a connection between the self-completeness  of gravity and a string theoretic completion.   For this  we refer the reader to \cite{dvali-gomez} and references therein.

\section*{Acknowledgements}
We wish to thank  Cesar Gomez, Gregory Gabadadze,  Oriol Pujolas, Slava Mukhanov,  Alberto Iglesias and Alex Pritzel for useful discussions. CG would also like to thank Leonard Susskind and Edward Witten for useful comments.
This work was supported in part by Humboldt Foundation under Alexander von Humboldt Professorship.  The work of  G.D. was also supported in part  by European Commission  under 
the ERC advanced grant 226371,  by  David and Lucile  Packard Foundation Fellowship for  Science and Engineering and  by the NSF grant PHY-0758032.

\appendix
\section{Corrections to energy-momentum tensor of the source}
\label{A}

The energy-momentum tensor of the colliding particles is modified during the scattering process due to its coupling to gravity. This modification is encoded in the conservation equation
\begin{equation}\label{cons}
\nabla_{\alpha}T^\alpha_\beta=0
\end{equation}
valid at all order. The conservation equation (\ref{cons}) is automatically obtained thanks to the diffeomorphism invariance of the action. At linear order, the conservation equation is obtained by the interaction
\begin{equation}\label{op}
\int d^4x \frac{h^{\alpha\beta}}{M_p} T^{(0)}_{\alpha\beta}\ ,
\end{equation}
where $T^{(0)}$ is the energy momentum tensor calculated in absence of gravity, or in other words, by considering the energy-momentum tensor as an {\it external} source. In fact by considering the linear diffeomorphic group under which the perturbation of the metric transform as $h_{\alpha\beta}=\partial_{(\alpha}\xi_{\beta)}$ we obtain the equation
\begin{equation}
\partial^\alpha T^{(0)}_{\alpha\beta}=0\ ,
\end{equation} 
which is the zeroth order in eq. (\ref{cons}). Obviously one may consider the first order in eq. (\ref{cons}). In this case the energy momentum tensor may no longer be consider as an external source. This is similar to radiative corrections in QED, for example. However, this contribution will only be important whenever the operator eq. (\ref{op}) will give a large contribution, i.e., after black hole formation (since the colliding particle masses are small with respect to the Planck scale). However this regime is hidden behind a Black Hole. Concluding, although it is true that, at full non-linear level, the energy-momentum tensor is not an external source, {\it it is} at linearized level, which is the regime considered in our paper.

The following computation is to show that  the first order corrections to the stress-energy tensor of the ``external'' particle are indeed negligible. In this case, the particle can no  longer be considered  as a point-like $\delta$-function source. Instead we model the particle as a perfect fluid ball of radius of the its Compton wavelength $R_c$ with constant density $\rho=\frac{M}{V}=const$ for $r<R_c=M^{-1}$, where $V$ is the volume of the ball and $M$ the mass of the particle. The stress-energy tensor of such a fluid ball is given by

\beq
T^{(0)}_{\alpha\beta}=(\rho+p)u_\alpha u_\beta+p g_{\alpha\beta} \; .
\eeq

We assume the matter to be non-relativistic to first approximation., i.e. $\rho\gg p$. In a static spacetime the fluid velocity 4-vector points in the same direction as the static Killing vector field: $u_\alpha\propto (dt)_\alpha$. Which in our coordinates means $u_\alpha\propto \delta^0_\alpha$. A timelike 4-velocity gives the constraint
\beq
u_\alpha u^\alpha=-1 \; ,
\eeq 
and it follows that $u_\alpha = \frac{1}{\sqrt{-g^{00}}}\delta^0_\alpha$.

On a Minkowski background, eq. (\ref{cons}) is satisfied by $\rho=const$, $p=0$ and $u_\alpha$ being a solution to the geodesic equation 
\beq
u_\alpha \nabla^\alpha u_\beta=0 \; .
\eeq
This source yields  the first order perturbations in the metric, see eq. (\ref{lineargraviton}). What is the effect of these perturbations on the source itself?

The 4-velocity up to first order corrections is 
\beq
u_\alpha= (1-\frac{1}{2}h^{00})\delta^0_\alpha \; .
\eeq

From eq. (\ref{lineargraviton}) we know that $h_{\mu\nu}=\frac{2 G m(r)}{r}\delta_{\mu\nu}$ with $m(r)=\int^{R_c}_0 d^3x \rho$.
Let us split eq. (\ref{cons}) in two orthogonal parts; one in the direction of $u_\alpha$ and the other orthogonal to it.
\begin{eqnarray}
u^\alpha\nabla_\alpha\rho + (\rho+p)\nabla_\alpha u^\alpha&=&0 \; , \label{udir}\\
(p+\rho)u^\alpha\nabla_\alpha u_\beta+(g_{\alpha\beta}+u_\alpha u_\beta)\nabla^\alpha p&=&0 \; . \label{uort}
\end{eqnarray}

Eq.  (\ref{udir}) gives $\partial_t\rho=0$ which is satisfied trivially. Eq. \ref{uort} gives us the correction to the pressure due to the selfinteraction of the gravitational source. In a static spacetime the pressure cannot depend on $t$ and we find 
\beq
-\frac{1}{2}\rho\, \partial_i h_{tt}+\partial_i p=0 \label{peq} \; ,
\eeq
where $i$ denotes the three spatial coordinates. Together with the boundary condition that $p(R_c)=0$ we find that $p^{(1)}(r)=\frac{1}{2}\rho( h^{(1)}_{00}(r)- h^{(1)}_{00}(R_c))$. So the first order correction to $T_{\alpha\beta}$ is given by

\beq
T_{\alpha\beta}^{(1)}=2 \rho u_\alpha^{(1)}u_\beta^{(0)}+p^{(1)}\eta_{\alpha\beta}\;,
\eeq
where $u_\alpha^{(1)}=-\frac{1}{2}h^{00}\delta^0_\alpha$ and $u_\beta^{(0)}=\delta^0_\beta$.

We see that the first order correction is always subleading as long as $h_{\mu\nu}\ll 1$. This is the point where a BH starts forming and hence our approximation ceases to be valid. 

It immediately follows that the gravitational binding energy, which is given by
\beq
E_B=M_p-M=\int^{R_c}_0\!\!\!d^3x\,h^i_i\rho \; ,
\eeq
where $M_p=\int^{R_c}_0 d^3x \sqrt{g^{(3)}}\rho$ is the proper mass defined by the proper volume integral over the density, is (as expected) negligible in the weak coupling regime, i.e. before the BH formation. This is also obviously true for any modification of gravity that does not violate energy conditions, or, in field theoretical terms, that do not at least propagate ghosts.

We conclude that we can safely neglect the back-reaction of the gravitational field on the source in the weak coupling regime.

\section{Strong Coupling Scale}
\label{B}

In this appendix we want to estimate the strong coupling scale $M_*$. The strong coupling scale is given by the minimal scale at which some scattering amplitudes become of order one. To put a bound on this scale, we consider a non-relativistic particle of mass $M$.

The gravitational potential can now be probed by an external static non-relativistic source $\tau_ {\mu\nu}= \delta_{\mu}^{0}\delta_{\nu}^{0} \, \delta^3(r-r') \, m$ and the strength of this interaction is set by the amplitude
\be
\label{eq:a1}
{\cal A}=\int_0^{\infty} \frac{h_{00}(r')}{M_P} \delta^3(r-r') \, m\, d^3r'=\frac{h_{00}(r)}{M_P}m \, .
\ee
Whenever ${\cal A}/m\sim 1$ unitarity is violated, as the probability of this process per unit time and probe mass is of order one.

For a given mass $M_v$, there is always a radius $r_v$ at which the unitarity bound of the theory is violated, or in other words
\be
2\int_0^\infty\!\!\!\!\!ds  \frac{\rho(s)}{M_P^2}\frac{e^{-\sqrt{s}r_v}}{r_v}M_v=1\ .\label{strongh}
\ee
In the previous expression we used the fact that one can always spectrally decompose $h_{00}(r)$ to be $\displaystyle{h_{00}(r)=2\int_0^\infty\!\!\!\!\!ds \frac{\rho(s)}{M_P}\frac{e^{-\sqrt{s}r}}{r}E}$, where all spin-2 and spin-0 poles were summed up.

By Heisenberg principle, $r_v$ cannot be smaller that the Compton wavelength of a particle with mass $M_v$, i.e. $r_v\geq M_v^{-1}$. Therefore, the minimal $M_v$ is obtained by inverting
\be
2\int_0^\infty\!\!\!\!\!ds \rho(s)e^{-\frac{\sqrt{s}}{M_v}}\left(\frac{M_v}{M_P}\right)^2=1\ ,
\ee
i.e.
\be
M_v=\frac{M_P}{\sqrt{I(M_v)}}\leq M_P\ .\label{sM}
\ee
The last inequality is a direct consequence of that
\be
I(M_v)\equiv2\int_0^\infty\!\!\!\!\!ds  \rho(s)e^{-\frac{\sqrt{s}}{M_v}}\geq 1\ .\label{noweaker}
\ee
(\ref{noweaker}), as explained in section (\ref{EGWG}), is related to the fact that any ghost free theory of gravity can only produce a stronger gravitational field than the one produced in the Einstein theory, for which $\rho(s)=\delta(s)$. By definition then the strong coupling scale of the theory is $M_*\leq M_v\leq M_P$.

A consistency check of (\ref{sM}) is obtained by considering that in Einstein gravity, in which $I=1$, the strong coupling scale is the Planck scale $M_P$.

\end{document}